\documentclass[structabstract]{aa}  
\usepackage{amstext,amsmath}
\usepackage{mathtools}
\usepackage{graphicx}
\usepackage{textcomp}
\usepackage{txfonts}
\usepackage{time}
\usepackage{subfig}
\usepackage{multicol}
\usepackage{color}
\usepackage{url}
\usepackage{pstricks} 
\usepackage{natbib,twoopt}
\usepackage[breaklinks=true,colorlinks=true,urlcolor=blue,citecolor=blue,linkcolor=blue]{hyperref} %% to avoid \citeads line fills
\bibpunct{(}{)}{;}{a}{}{,} %% natbib format for A&A and ApJ
\makeatletter
\newcommandtwoopt{\citeads}[3][][]{\href{http://adsabs.harvard.edu/abs/#3}%
{\def\hyper@linkstart##1##2{}%
\let\hyper@linkend\@empty\citealp[#1][#2]{#3}}}
\newcommandtwoopt{\citepads}[3][][]{\href{http://adsabs.harvard.edu/abs/#3}%
{\def\hyper@linkstart##1##2{}%
\let\hyper@linkend\@empty\citep[#1][#2]{#3}}}
\newcommandtwoopt{\citetads}[3][][]{\href{http://adsabs.harvard.edu/abs/#3}%
{\def\hyper@linkstart##1##2{}%
\let\hyper@linkend\@empty\citet[#1][#2]{#3}}}
\newcommandtwoopt{\citeyearads}[3][][]%
{\href{http://adsabs.harvard.edu/abs/#3}
{\def\hyper@linkstart##1##2{}%
\let\hyper@linkend\@empty\citeyear[#1][#2]{#3}}}
\makeatother

\usepackage{url}
\usepackage{epic}
\usepackage{rotating}

\usepackage{multirow} 
\usepackage{array} 
\usepackage{balance}
\usepackage{epic}
\usepackage{xcolor}
\definecolor{dark-gray}{gray}{0.4}

\usepackage{xifthen}

\newcommand{\hed}{He~\textsc{i} D\textsubscript{3} }
\newcommand{\hedc}{He~\textsc{i} D\textsubscript{3}}
\newcommand{\hei}{He~\textsc{i} 10830~\AA\;}
\newcommand{\heic}{He~\textsc{i} 10830~\AA}
\newcommand{\der}{{\rm d}}

\makeatletter
\def\url@leostyle{%
  \@ifundefined{selectfont}{\def\UrlFont{\sf}}{\def\UrlFont{\tiny\ttfamily}}}
\makeatother
\begin{document}
\authorrunning{T. Libbrecht et al. }

\title{Line formation of He I D{\small 3} and He I 10830 \AA\;in a small-scale reconnection event}
%\subtitle{bla}

%\author{Tine Libbrecht\inst{1} \and Johan Pir\'es Bj\o rgen\inst{1} \and Jorrit Leenaarts\inst{1} \and Jaime de la Cruz Rodr\'iguez\inst{1} \and Viggo Hansteen\inst{2} \and Jayant Joshi\inst{2}}
\author{Tine Libbrecht \inst{1} \and Johan P. Bj\o rgen\inst{1,2} \and Jorrit Leenaarts\inst{1} \and Jaime de la Cruz Rodr\'iguez\inst{1} \and Viggo Hansteen\inst{2,3,4,5} \and Jayant Joshi\inst{2}}

\institute{Institute for Solar Physics, Dept. of Astronomy, Stockholm University, Albanova University Center, SE-10691 Stockholm, Sweden \email{tine.libbrecht@astro.su.se}
\and Rosseland Centre for Solar Physics, University of Oslo, PB 1029 Blindern, 0315 Oslo, Norway
\and Institute of Theoretical Astrophysics, University of Oslo, P.O. Box 1029 Blindern, N-0315 Oslo, Norway
\and Lockheed Martin Solar \& Astrophysics Laboratory, 3251 Hanover St., Palo Alto, CA 94304, USA
\and Bay Area Environmental Research Institute, NASA Research Park, Moffett Field, CA 94035, US }

% \date{Received September 15, 1996; Accepted March 16, 1997}
\date{Draft: \now\ \today}
\frenchspacing

\abstract
{Ellerman bombs (EBs) and UV bursts are small-scale reconnection events occurring in the upper photosphere to the chromosphere. Recently, \citetads{2017A&A...598A..33L} discovered that these events can have emission signatures in the He \textsc{i} D3 and He \textsc{i} 10830 \AA\ lines, suggesting that their temperatures are higher than previously expected.}
{We aim to explain line formation of \hed and \hei in small-scale reconnection events.}
{We make use of a simulated EB, present in a Bifrost-generated radiative Magnetohydrodynamics (rMHD) snapshot. The resulting \hed and \hei line intensities are synthesized in 3D using the non-LTE Multi3D code. The presence of coronal EUV radiation is included self-consistently. We compare the synthetic helium spectra with observed SST/TRIPPEL raster scans of EBs in \hei and \hedc.}
{Emission in \hed and \hei is formed in a thin shell around the EB at a height of $\sim 0.8$ Mm while the \hed absorption is formed above the EB at $\sim 4$ Mm. The height at which the emission is formed corresponds to the lower boundary of the EB, where the temperature increases rapidly from $6\cdot 10^3$ K to $10^6$ K. The synthetic line profiles at a heliocentric angle of $\mu=0.27$ are qualitatively similar to the observed ones at the same $\mu$-angle when it comes to dynamics, broadening and line shape: emission in the wing and absorption in the line core. The opacity in \hed and \hei is generated via photoionization-recombination driven by EUV radiation that is locally generated in the EB at temperatures in the range of $2\cdot 10^4 - 2\cdot 10^6$ K and electron densities between $10^{11}$ and $10^{13}$ cm\textsuperscript{--3}. The synthetic emission signals are a result of coupling to local conditions in a thin shell around the EB, with temperatures between $7\cdot 10^3$ and $10^4$ K and electron densities ranging from $\sim 10^{12}$ to $10^{13}$ cm\textsuperscript{--3}. Hence, both strong non-LTE as well as thermal processes play a role in the formation of \hed and \hei in the synthetic EB/UV burst that we studied.}
{In conclusion, the synthetic \hed and \hei emission signatures are an indicator of temperatures of at least $2\cdot 10^4$ K and in this case as high as $\sim 10^6$ K.}

\maketitle

\section{Introduction}\label{sec:intro}
Both the \hed and \hei lines originate from transitions between states in the triplet system of He \textsc{i}. Neutral helium consist of a singlet system (quantum number $S=1$) and a triplet system (quantum number $S=3$) between which radiative electric dipole transitions are forbidden. Since \citetads{1939ApJ....89..673G} discovered that the transitions in the triplet system of neutral helium are anomalously bright compared to the singlet transitions, the population mechanism of the helium triplet levels has been debated. Many modelling studies have suggested a photoionization-recombination mechanism (PRM) in which the required EUV photons are originating in the corona and impinging on the chromosphere where neutral helium gets ionized and then recombines into both the singlet and the triplet states \citepads{1975ApJ...199L..63Z,1997ApJ...489..375A,2008ApJ...677..742C,2016A&A...594A.104L}

%In subsequent decades, the helium line formation discussion focused mainly on whether the EUV-flux from the corona is sufficient for the PRM to take place. Additionally, it was noted that images of \hei exhibit sub-arcsecond spatial structuring \citepads{1985SoPh...97...35L}. The latter is in apparent contradiction with a diffuse EUV-radiation field originating in the  corona. 
\citetads{2016A&A...594A.104L} used a 3D rMHD simulation and 3D non-LTE spectral synthesis of the \hei line to show that the source of the ionizing photons is not only located in the $10^6$ K corona but also in the transition region where ionizing photons are originating at $T\sim 8\cdot 10^4$ K. The ionizing photons at this temperature are emitted in locations that are spatially very close to the upper chromosphere where \hei and \hed are formed, which is how sub-arcsecond structure in \hei and \hei images is formed.

All findings of \citetads{2016A&A...594A.104L} are however derived from a quiet-sun-like rMHD snapshot. Line formation of \hei in active regions and in reconnection targets such as flares is still unknown, especially the role of electron collisions in comparison to the PRM in flares (see e.g. \citeads{1992ApJ...386..364L,2005A&A...432..699D,2014ApJ...793...87Z,2015ApJ...814..100J,2016ApJ...819...89X} for \hei and \citeads{2013ApJ...774...60L,2019A&A...621A..35L} for \hedc).

Ellerman bombs (EBs) and UV bursts are examples of reconnection events on smaller scales that have attracted a lot of attention in recent years, specifically with regard to their temperatures. The most characteristic spectral signature of EBs are their moustache shaped line profiles in H${\alpha}$, which have been observed for over a 100 years \citepads{1917ApJ....46..298E}. However, with the launch of the Interface Region Imaging Spectrometer (IRIS, \citeads{2014SoPh..289.2733D}), it was discovered that EBs can exhibit strongly enhanced and broadened emission profiles in the Si \textsc{iv} 1400 \AA\ doublet (e.g. \citeads{2014Sci...346C.315P,2015ApJ...812...11V,2016ApJ...824...96T,2017A&A...598A..33L,2019A&A...627A.101V,2019ApJ...875L..30C,2020A&A...633A..58O}). Under the assumption of coronal equilibrium, the Si \textsc{iv} 1400 \AA\ doublet has a formation temperature of $\sim 8\cdot 10^4$ K which is not compatible with EB temperatures of $T\leq 10^4$ K as calculated via semi-empirical modelling (e.g.~\citeads{2013A&A...557A.102B,2014A&A...567A.110B,2017RAA....17...31F}).

\citetads{2017A&A...598A..33L} presented spectral raster scans of EBs observed with TRIPPEL at the SST of \hedc, \hei and H$\beta$ lines in co-observation with IRIS. It was found that EBs/UV bursts can have emission signals in \hed and \heic, which the authors have interpreted as evidence for EBs having temperatures ranging between  $T\sim 2\cdot 10^4-10^5$ K. 

In this paper, we aim to dig deeper into the line formation of \hed and \hei in these type of events, with the goal of understanding under which conditions \hed and \hei emission can be generated in EBs. Therefore, we make use of a simulated EB generated with the rMHD code Bifrost. Radiative MHD simulations have been used before to study spectral diagnostics of EBs and UV bursts \citepads{2013ApJ...779..125N,2017A&A...601A.122D,2017ApJ...839...22H,2017ApJS..229....5D}. Those studies demonstrated that the spectral diagnostics of EBs and UV bursts are reasonably well-reproduced during the reconnection events found in the rMHD simulations. The details of the 3D rMHD simulation used and the event studied in this paper are described in \cite{Hansteen}, while we focus on helium line formation in the EB. Our results are possibly also relevant for flares or other events causing emission in \hed and \heic, for example shocks \citepads{2007A&A...462.1147L}.

\section{Method}\label{method}
\subsection{rMHD simulation and event description}
We made use of an rMHD simulation with the Bifrost code \citepads{2011A&A...531A.154G} which was run and described in detail by \cite{Hansteen}. The MHD equations are solved with the assumption of LTE ionization for the plasma. Radiative transfer was included in the energy equation via approximate terms describing the photospheric and chromospheric optically thick radiative losses and the coronal optically thin radiative losses of the dominant spectral lines \citepads{2012A&A...539A..39C}. The absorption of coronal radiation in the chromosphere has been treated in 1D, meaning that the coronal radiative losses propagate vertically. 

The numerical domain spans 24 Mm horizontally and 17 Mm vertically from 2.5 Mm below the photosphere up to 14.5 Mm into the corona. The simulation was run with $768\times 768 \times 768$ grid points with uneven vertical sampling so that the sampling is dense when the scale height is small.

The Bifrost simulation was run with the aim of studying a flux emergence region and reconnection events. The original model is an enhanced network simulation with small-scale magnetic field patches of up to 1500 G. This model is partially based on the publicly available model \citepads{2016A&A...585A...4C} but using higher resolution and an LTE equation of state. A magnetic sheet with a horizontal magnetic field strength of $B_y=2000$ G is added at the bottom boundary in the convection zone. The magnetic sheet rises via a combination of convective motions and buoyancy and emerges into the photosphere in small-scale elements. At this point, the field only continues to rise in locations where it is sufficiently strong and where the field is not trapped by dense material. These conditions give rise to the presence of undulatory field lines where oppositely directed magnetic fields are pushed closely together by oppositely directed photospheric flows. In such locations, a current sheet forms and reconnection happens in the form of EBs and UV bursts. 

\begin{figure*}
\includegraphics[scale=1]{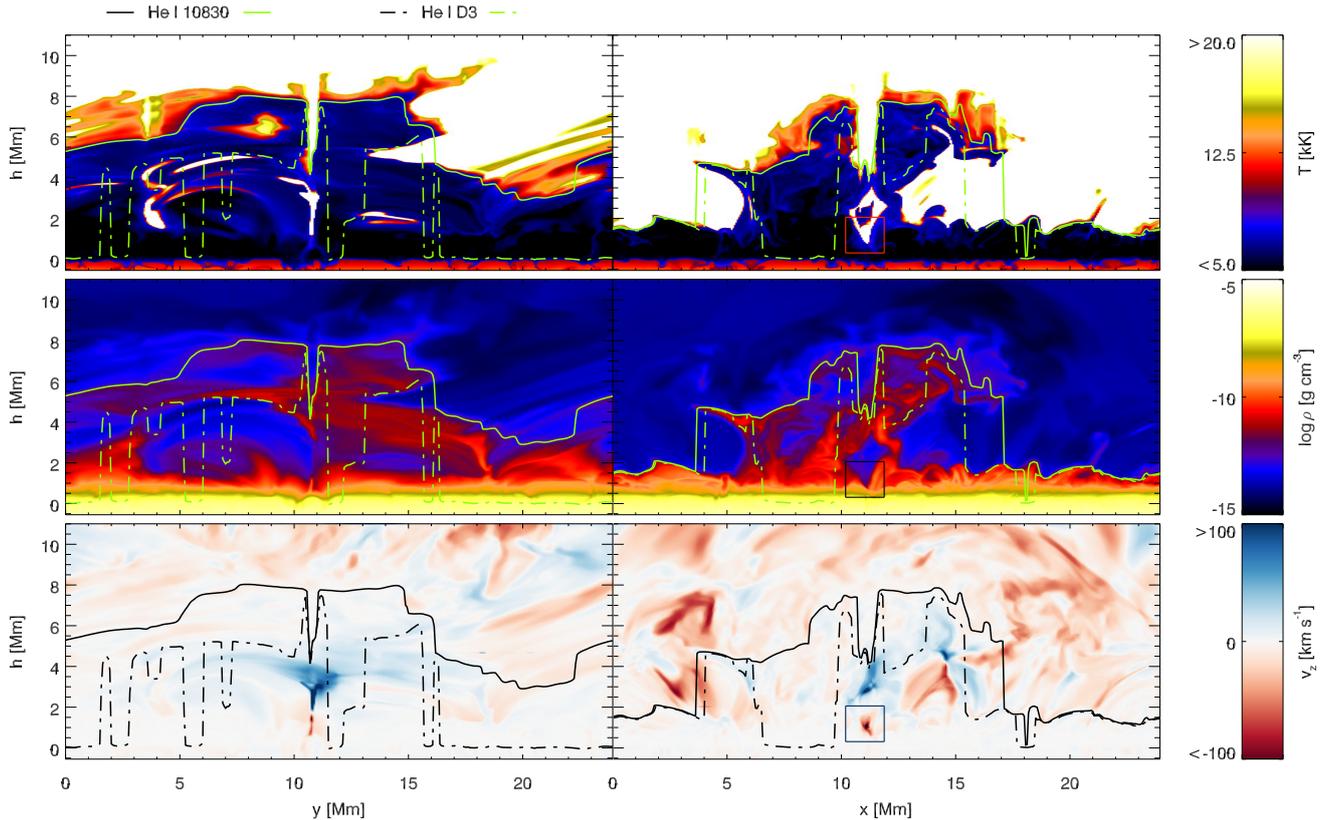}
\caption{Vertical cuts through the rMHD snapshot intersecting the EB location. The left column is a cut along the $y$-axis, and the right column is a cut along the $x$-axis. We display the temperature, density (on a logarithmic scale) and the vertical velocity together with $h_{\max}$ (see Eq.~\ref{hmax}) of \hei (solid line) and \hed (dash dotted line). The red and black boxes in the right panels indicate the region that we zoom into in Figs.~\ref{pars}, \ref{radpans} and \ref{destr}. \label{vertical_cuts}}
\end{figure*}

In the simulation, at least one such event is present in the middle of the domain at $x=11$, $y=11$ Mm. In this paper, we have selected a snapshot in which the event reaches a temperatures of up to $\sim 3$ MK between at a height of $0.8-3.5$ Mm above the photosphere. It has been classified as both an EB and UV burst by \cite{Hansteen} because of the nominal moustache shape of the H$\alpha$ spectral profile and the broadened and enhanced Si \textsc{iv} 1400 \AA\ doublet. The presence of both these spectral signatures is an incentive for us to use this snapshot, since both of these were present in the EBs observed by \citetads{2017A&A...598A..33L}. Vertical magnetic field of opposite polarity are colliding, with a vertical field strength of $\sim 600$ G at 0.15 Mm. Bi-directional jets are present with vertical velocities of the order of $\sim 100$ km s\textsuperscript{-1}. A detailed description of the development of the current sheet and the event evolution is given in \cite{Hansteen}. Vertical cuts along the x- and y-axis through the EB are shown in Fig~\ref{vertical_cuts} for temperature, density and vertical velocity. 

\subsection{Multi3D and the model atom}\label{modelatom}
To synthesize the \hed and \hei spectral lines, we made use of the 3D non-LTE radiative transfer code Multi3D \citepads{2009ASPC..415...87L}. The latter solves the statistical equilibrium equations using multi-level accelerated $\Lambda$-iteration (MALI, \citeads{1991A&A...245..171R,1992A&A...262..209R}). We use an A4 angle quadrature set consisting of 24 angles. All background opacities including hydrogen are calculated with the Uppsala opacity package \citepads{Gustafsson1973}. We used only every second grid point in the $x$ and $y$ direction from the rMHD snapshot for our radiative transfer calculations and the top boundary has been clipped, so the input cube is reduced to dimensions $384\times 384\times 606$.

The helium model atom is a simplified 16-level version of the 33-level model atom used by \citet{2014ApJ...784...30G}, where the sources of the atomic data are listed in detail. It contains 12 He \textsc{i} levels, 3 He \textsc{ii} levels and the He \textsc{iii} continuum. In Fig.~\ref{modelatom}, we show the details of the included energy levels and transitions in He~\textsc{i}.
\begin{figure}
\includegraphics{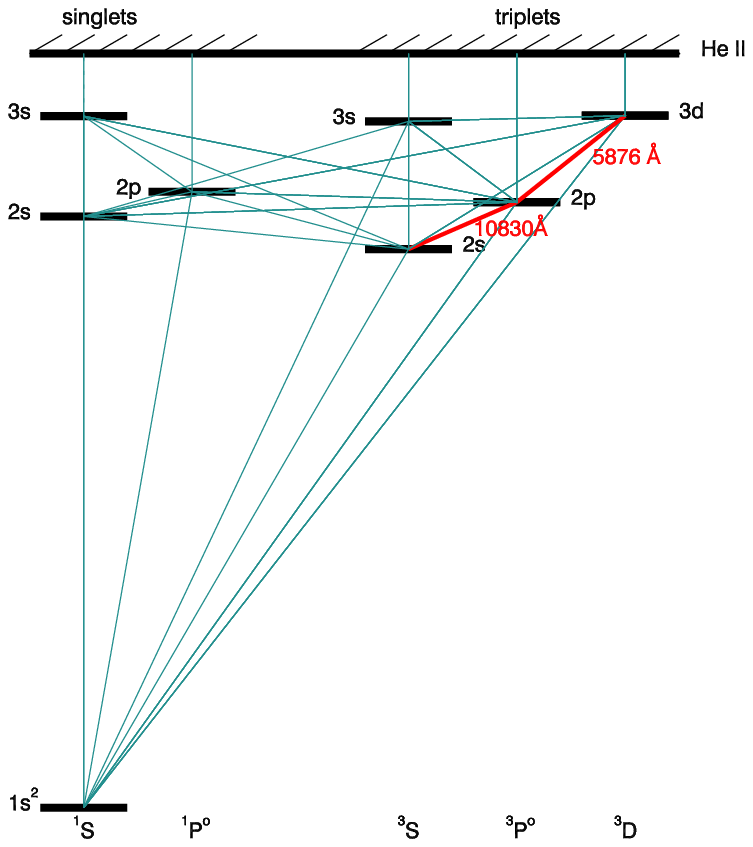}
\caption{Schematic view of the model atom for He~\textsc{i}. The blue lines represent the transitions present in the model atom file, both collisional and radiative. The red lines indicate the \hei and \hed spectral lines. Energy levels $1s 2p\,^3P_{0,1,2}$ and $3d\,^3D_{1,2,3}$ are each depicted in the diagram as one energy level but actually consist of three each. The total number of levels in the model atom is 16 of which 12 for He~\textsc{i}.\label{modelatom}}
\end{figure}
Our primary goal is to study the \hed and \hei spectral lines which are part of the triplet system. Therefore, all seven levels that make up these transitions are included unaltered: $1s 2s\,^3S_1$, $1s 2p\,^3P_{0,1,2}$ and $1s 3d\,^3D_{1,2,3}$ while the $1s 3s\,^3S_2$ level and the $1s 3p\,^3P_{0,1,2}$ levels have been merged. In the singlet system, we include the ground level $1s^2\,^1S_0$, and the levels $1s 2s\,^1S_0$ and $1s 2p\,^1P_1$. The levels $1s 3s\,^1S_0$ and $1s 3p\,^1P_1$ have been merged. All higher excited states of neutral helium are not used. In the He \textsc{ii} atom, only the He \textsc{ii} continuum level has been kept unchanged. The levels $2p\,^2P_{\frac{1}{2},\frac{3}{2}}$ and $2s\,^2S_\frac{1}{2}$ are merged, as well as the levels $3s\,^2S_\frac{1}{2}$, $3p\,^2P_{\frac{1}{2},\frac{3}{2}}$, and $3d\,^2D_{\frac{3}{2},\frac{5}{2}}$.

We have also added collisional transitions to the model atom that were missing in the original 33-level model atom. The most important missing collisional transitions were those between the energetically close $J$-splitted sub-levels of the \hei and \hed levels themselves (i.e. for example $^3P_0$ to $^3P_1$), because these collisional rates are roughly 10 orders of magnitude larger than the collisional rates between for example $^3P_0$ to $^3D_1$. The rates were obtained via the Van Regemorter prescription for collisional excitation \citepads{1962ApJ...136..906V}.

The helium model atom includes 26 bound-bound transitions, 62 collisional transitions, and 15 bound-free transitions. The frequency grid contains 1139 points of which the \hei and \hed transitions are sampled with 278 frequency points each. Of these 278 frequency points, 226 are sampling the lines equidistantly between $-224$ and $+ 224$ km s\textsuperscript{--1}, reaching the largest chromospheric velocities in the rMHD snapshot.

Coronal radiation is included in the Multi3D calculations as described in detail by \citetads{2016A&A...594A.104L}. The coronal emissivity is defined as 
\begin{equation}\label{emisdens}
\psi_\nu =\Lambda_\nu(T)n_{\rm e}n_{\rm H},
\end{equation} with $n_{\rm e}$ the electron number density, $n_{\rm H}$ the hydrogen number density and $\Lambda_\nu(T)$ the coronal emissivity per electron and per hydrogen atom as a function of temperature. $\Lambda_\nu(T)$ is calculated from CHIANTI \citepads{2009A&A...498..915D} with the assumption of coronal equilibrium ionization for all elements with lines present in the EUV spectrum shortward of the ionization edge of He \textsc{i} at 504 \AA. These calculations do not include hydrogen and helium lines since these are already accounted for in Multi3D as background source and active element respectively. The coronal emissivity $\psi_\nu$ calculated via CHIANTI is then remapped onto the coarse frequency grid used by Multi3D while preserving the frequency-integrated emissivity.

\subsection{Observational data}
The \hei and \hed spectra were obtained via raster scans with the TRI-Port Polarimetric Echelle-Littrow spectrograph (TRIPPEL, \citeads{2011A&A...535A..14K}) at the Swedish 1-m Solar Telescope (SST, \citeads{2003SPIE.4853..341S}) on 2015-08-01 between 07:51 UT and 10:10 UT. The data acquisition and reduction is described in detail in \citetads{2017A&A...598A..33L}.

\section{Results}
\subsection{Helium spectra across the snapshot domain}
\begin{figure*}
\includegraphics[scale=1]{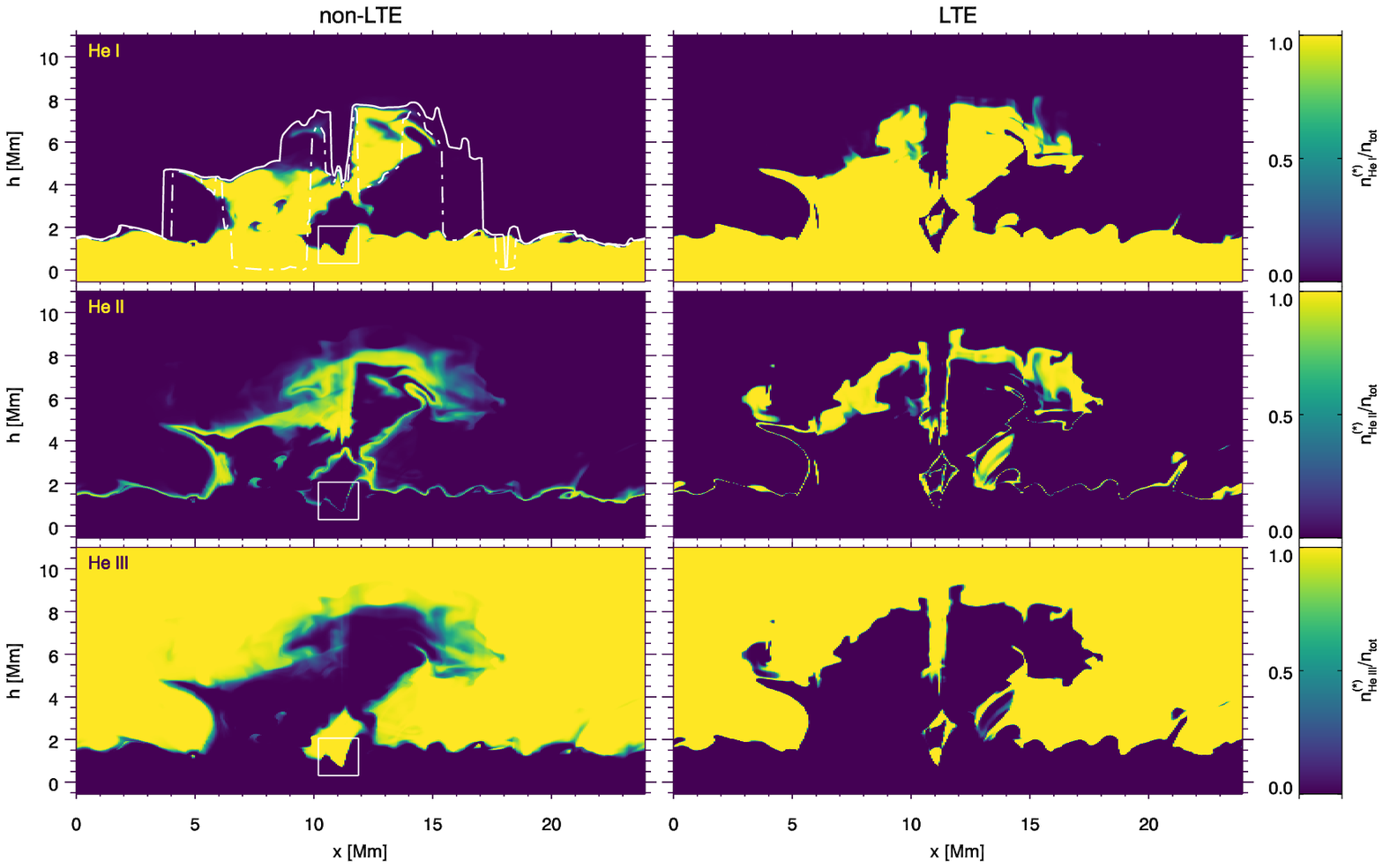}
\caption{Vertical cuts through the rMHD snapshot intersecting x-axis at the EB location. The left column is the non-LTE He \textsc{i}, \textsc{ii}, and \textsc{iii} population relative to the total population. The right column are the LTE He \textsc{i}, \textsc{ii}, and \textsc{iii} populations relative to the total population. $h_{\max}$ (see Eq.~\ref{hmax}) of \hei (solid line) and \hed (dash dot line) are plotted in the upper left panel. The white boxes in the left panels indicate the region that we zoom into in Figs.~\ref{pars}, \ref{radpans} and \ref{destr}. \label{ion_cuts}}
\end{figure*}

\begin{figure*}
\includegraphics[scale=1]{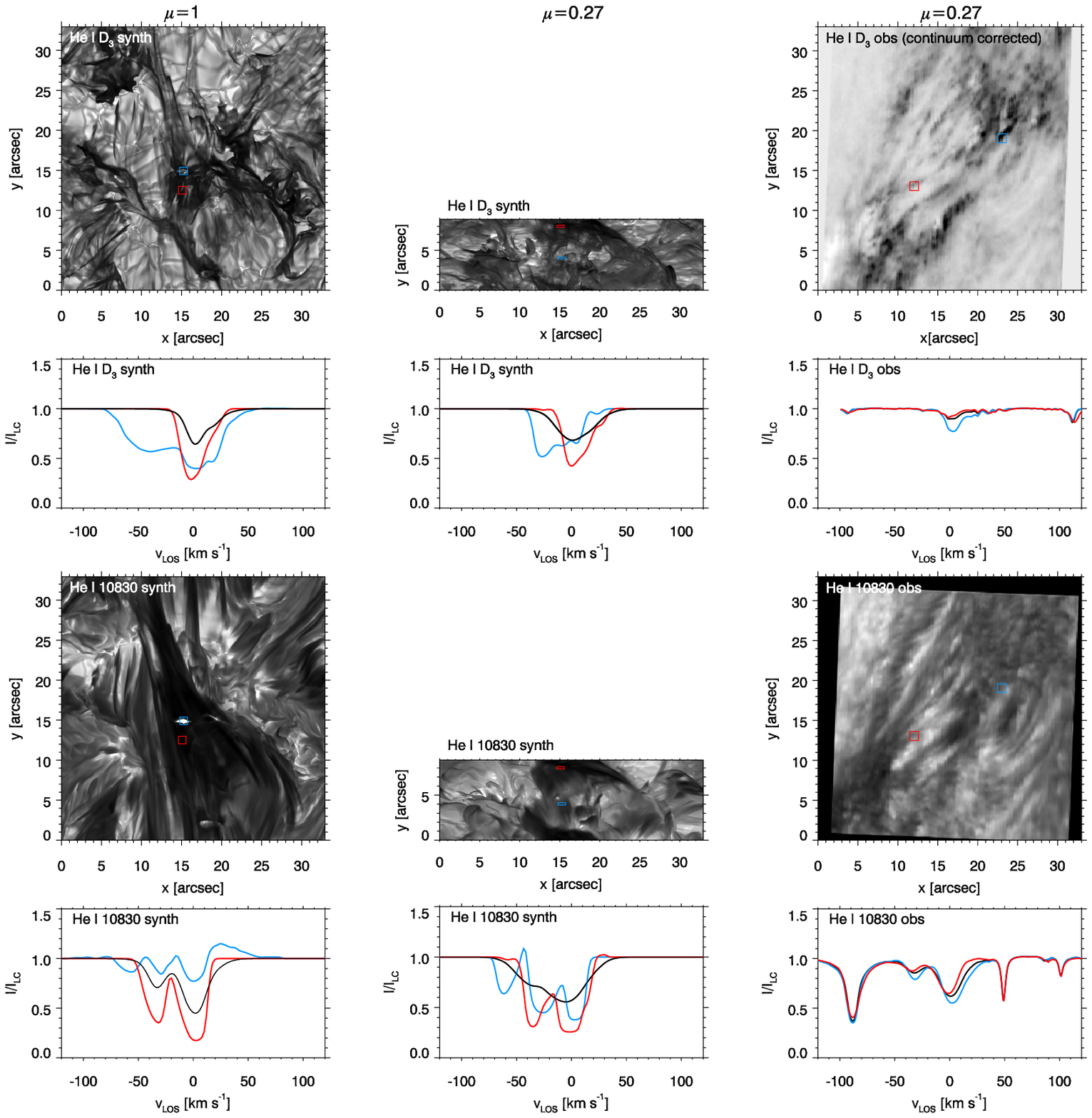}
\caption{A comparison between a simulated and observed flux emergence region in \hei and \hedc. The top two rows show \hed images and spectra while the lower two rows show \hei images and spectra. The left column is synthetic data at $\mu=1$, the middle column is synthetic data at $\mu=0.27$ and the right column is observed data at $\mu=0.27$. The red and blue squares on the images indicate areas over which the spectra is averaged that is displayed in the corresponding color on the panel below the image. The black spectrum is an average over the entire field of view shown in the image. \label{show_ie}}
\end{figure*}

Before focusing on the EB, we take a brief look at the entire snapshot domain. Vertical cuts of the rMHD snapshot are shown in Fig.~\ref{vertical_cuts} where the EB is present in both columns at $x=11$, $y=11$ Mm. In the EB location, the temperature rises steeply to coronal values of the order of $10^5-10^6$ K, the density drops, and bi-directional jets are present with velocities of the order of 100 km s \textsuperscript{--1}. 

In Fig.~\ref{vertical_cuts}, we display $h_{\max}$ of \hei and \hedc , defined as 
\begin{equation}\label{hmax}
h_{\max}=\max\left[h_{\nu}(\tau_{\nu}=1)\right].
\end{equation}
This usually corresponds to the height where the line core is formed. For \heic , $h_{\max}$ is situated everywhere at the interface between the chromosphere and the transition region, where the temperature rises steeply to coronal temperatures. In this snapshot, \hei is optically thick almost everywhere. The \hed line is generally formed at lower heights compared to \hei since it has lower opacity. The \hed absorption fluctuates between being optically thick and and optically thin, the latter is obvious in locations where the $h_{\max}$ of \hed drops suddenly down to the photosphere. The vertical cuts in Fig.~\ref{vertical_cuts} demonstrate that flux emergence in the simulation has pushed the corona upwards in a large part of the domain towards heights of 6-8 Mm. Regions that display a more nominal temperature stratification are e.g. $x=0$-$4$ Mm and $x=18$-$24$ Mm and we see that in these regions, the formation height of \hed and \hei coincide.

Even though Bifrost treats ionization in LTE in this simulation, with its implications to the heat budget, Multi3D takes into account non-LTE ionization of helium with the assumption of statistical equilibrium. The difference between non-LTE and LTE helium ionization is shown in Fig.~\ref{ion_cuts}. The non-LTE case takes into account radiation, making ionization non-local. Therefore, the He \textsc{ii} region is more diffuse than in the LTE case where it simply follows the steep temperature stratification. The opacity in the He \textsc{i} triplet system scales with the He \textsc{ii} population, since the triplet system is generally populated via recombination cascades from the He \textsc{ii} continuum (PRM). Sufficient He \textsc{i} has to be available at the same time as He \textsc{ii} in order to populate the He \textsc{i} triplet levels. The vertical cuts demonstrate that the non-LTE case is more likely to have substantial He \textsc{ii} and He \textsc{i} co-existing at the same locations. It is hence important to take into account non-LTE helium ionization to obtain (more) realistic \hei and \hed intensities. %\citetads{2014ApJ...784...30G} have used time dependent modelling in 1D to show that even statistical equilibrium is not valid under dynamic conditions and full time-dependent modelling should be done to get the correct helium intensities. This is however not feasible yet in 3D.

The emergent line core intensities of \hei and \hed are shown in Fig.~\ref{show_ie} and compared with observations of a flux emergence region in \hei and \hedc (same observations discussed as in \citeads{2017A&A...598A..33L}). Both the synthetic and observed \hei image show long and thick fibrils and a thick canopy structure, while the less active areas have a more grainy appearance. The average synthetic \hei absorption equals $0.6 I_{\rm LC}$, where $I_{\rm LC}$ is the local continuum intensity at $\mu=0.27$. This value is similar to the observed average absorption at the same $\mu$-angle. 

For the \hed line the difference is more substantial: the observations have to be continuum corrected to even notice the \hed absorption while the synthetic absorption gets optically thick in many locations and is very obvious over almost the entire domain. The absorption in the average synthetic \hed profile reaches $0.7 I_{\rm LC}$ while the average observed absorption is $0.9 I_{\rm LC}$ for $\mu=0.27$. 

The rMHD snapshot harbours velocities of the order of 100 km $\rm s^{-1}$ at the \hei and \hed formation height, while such high velocities are very rare in \hed and \hei observations of flux emergence regions. Those extreme velocities are causing the synthetic line profiles to exhibit multiple velocity components and hence a more complicated appearance, compared to the observed profiles.

\begin{figure*}[h]
\includegraphics[scale=1]{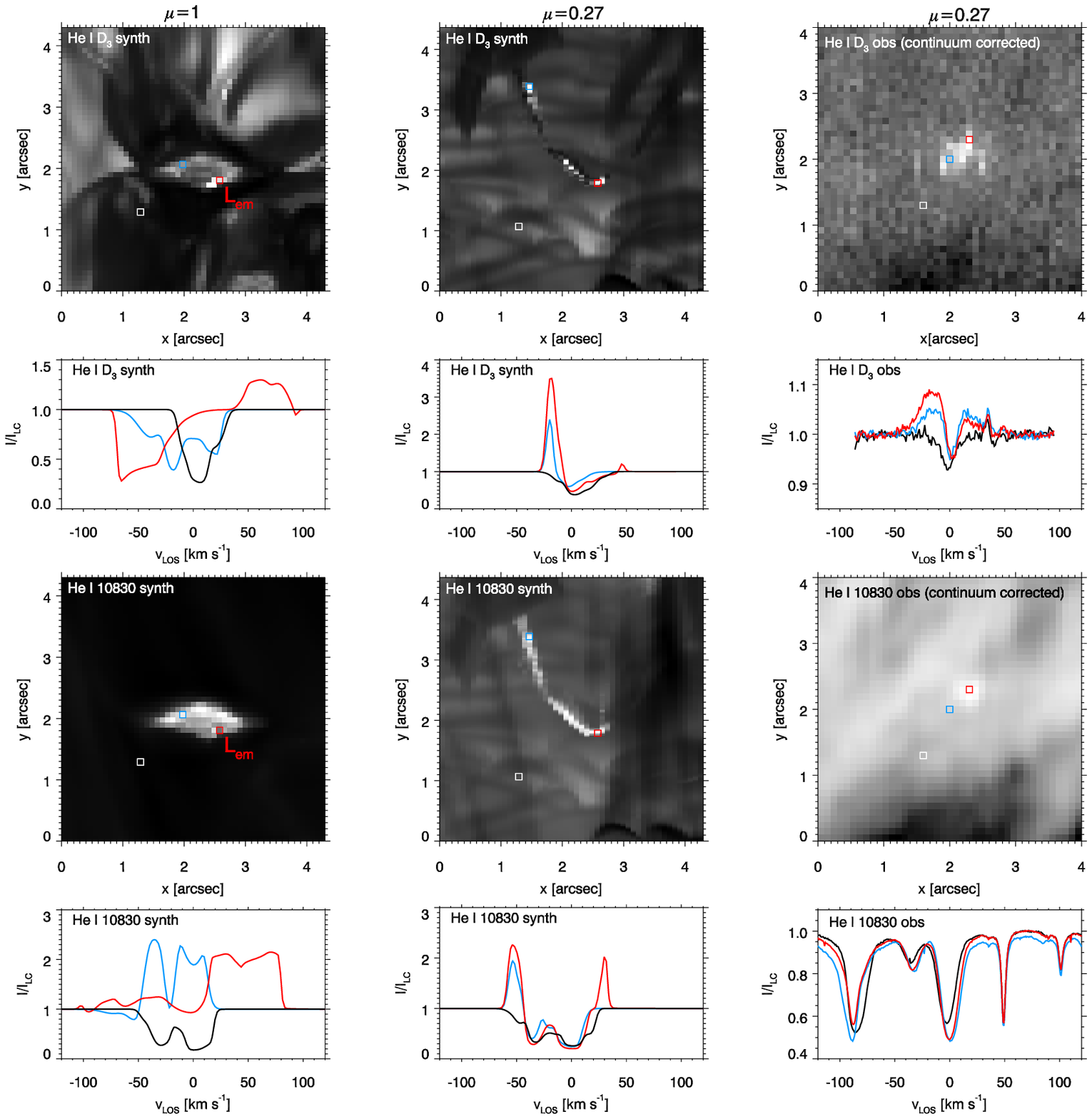}
\caption{A comparison between a simulated and an observed EB in \hei and \hedc. The top two rows show \hed images and spectra while the lower two rows show \hei images and spectra. The left column is synthetic data at $\mu=1$, the middle column is synthetic data at $\mu=0.27$, and the right column is observed data at $\mu=0.27$. The red and blue squares on the images indicate the area over which the spectra is averaged that is displayed in the corresponding color on the panel below the image (the white square corresponds to the black profile). The label $\rm L_{em}$ indicates the pixel with maximum emission in \hedc, which we study in detail in Sec.~\ref{lineform}. The black spectrum is an average over the entire field of view shown in the image.  The emission feature in the observed \hed spectrum at $v_{\rm LOS}\sim 40\;\rm km\; s^{-1}$ is an artifact resulting from a telluric correction applied to the spectra. Details of this procedure can be found in \citetads{2017A&A...598A..33L}.\label{obs_synth_zoom}}
\end{figure*}

\subsection{EB helium profiles}

Figure \ref{obs_synth_zoom} shows a zoom on an EB in both the synthetic and the observed data and the \hed and \hei spectra. First, we concentrate on the synthetic images and spectra at $\mu=1$. The EB has a slightly elongated appearance and is bright in both the line core image of \hed and \heic. The EB has a \hei line profile in pure emission while most \hed profiles are pure absorption profiles. However, some of the synthetic \hed profiles show a combination of emission and absorption, with the emission redshifted towards with a velocities of $\sim 70$ km $\rm s^{-1}$. 

The appearance of the EB changes drastically when the outgoing intensity is calculated along highly inclined rays. At an angle of $\mu=0.27$, the EB is one-pixel thin and very elongated in both \hei and \hed with a length of more than one arcsec. At this viewing angle, all synthetic \hed and \hei EB profiles are a combination of emission in the line wing and absorption in the line core. This demonstrates that the combination of obstruction by fibrils and the line-of-sight velocity are the dominant properties that give the helium EB profiles their shape.

The synthetic \hed and \hei EB profiles at $\mu=0.27$ match the observed profiles on a qualitative level: occurrence of emission in the wing, absorption in the core, and the dynamics seems to be largely reproduced. Large blue- and/or redshifts are present, depending on the viewing angle. Also, the broadening is of the same order of magnitude in the observed and the synthetic profiles. However, the synthetic emission features are much stronger compared to the observed ones. Also, the synthetic \hei emission is more intense than the synthetic \hed emission, while in the observations, the \hei is barely visible and the \hed emission is more prominent. However, the maximum observed \hed emission is still only at $\sim$ 1.1$I_{\rm LC}$ while the synthetic \hed emission is at $\sim$ 3.5 $I_{\rm LC}$ when comparing at equal viewing angles of $\mu = 0.27$. 

Despite the differences between our observations and the synthetic profiles, the 3D rMHD simulation provides a remarkable chance to study line formation in detail, which cannot be done with the observations. The fact that there is emission present in the synthetic EB profiles and that there is some qualitative overlap between the observations and simulation is encouraging, and we aim to pinpoint the cause of the synthetic emission. Therefore, we have selected the profile with maximum \hed emission to study in more detail. We name $\rm L_{em}$ the location at which this profile is located.
\begin{figure}
\includegraphics[scale=1]{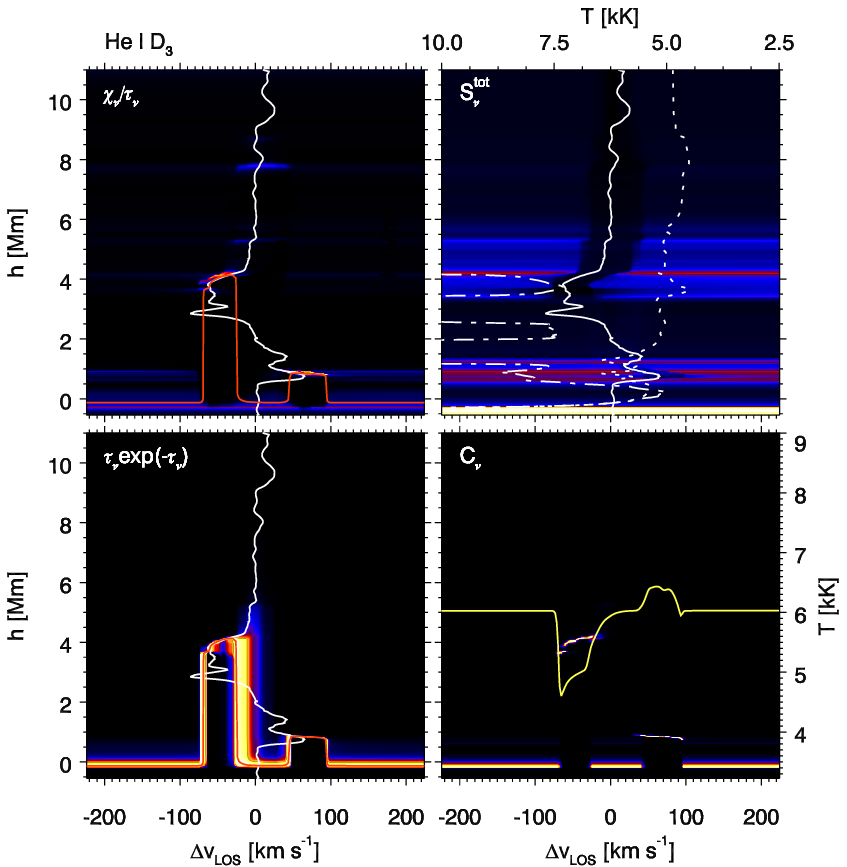}
\caption{Line formation plot with all factors that make up the contribution function (Eq.~\ref{contr}) of the EB \hed emission profile at location $\rm L_{em}$ (see Fig.~\ref{obs_synth_zoom}) and at a heliocentric viewing angle of $\mu =1$. Top left: opacity $\chi_\nu$ divided by optical depth $\tau_\nu$. The $h_{\max}$ surface (Eq.~\ref{hmax}) is shown as a solid red line and the vertical velocity $v_z$ as a white solid line. Top right: the total source function $S^{\rm tot}_\nu$ (Eq.~\ref{sf}). The vertical velocity $v_z$ is shown as a white solid line, the temperature $T$ as a dot-dash line and the line source function $S^l$ in temperature units as a dotted line. Bottom left: $\tau_\nu e^{-\tau_\nu}$ which displays the height range where $\tau_\nu$ is close to 1. The $\tau_\nu=1$ surface is shown as a solid red line and the vertical velocity $v_z$ as a white solid line. Bottom right: the total contribution function. The emergent intensity is shown on a temperature scale as a yellow solid line. \label{fourpanel}}
\end{figure}

\begin{figure}
\includegraphics[scale=1]{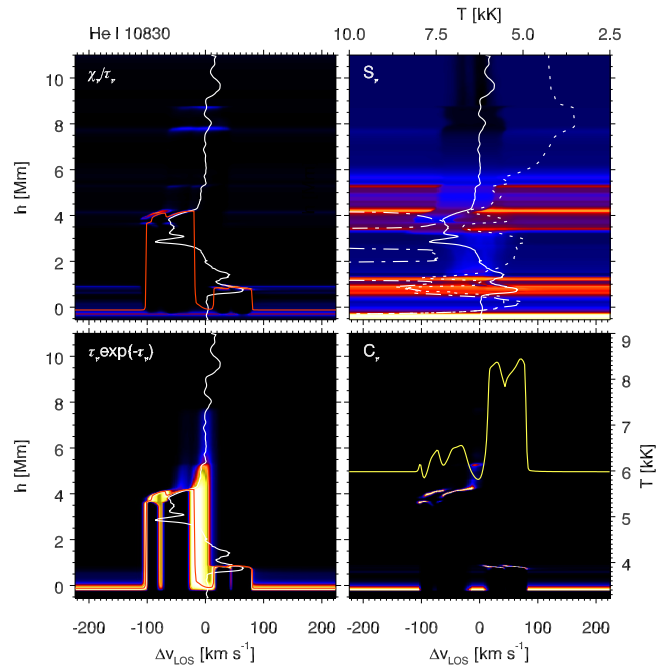}
\caption{Same as Fig.~\ref{fourpanel} but for \heic: line formation plot with all factors that make up the contribution function (Eq.~\ref{contr}) of the EB \hei emission profile at location $\rm L_{em}$ (see Fig.~\ref{obs_synth_zoom}) and at a heliocentric viewing angle of $\mu =1$. \label{fourpanelhei}}
\end{figure}

\subsection{EB helium line formation} \label{lineform}
\begin{figure}
\includegraphics[scale=1]{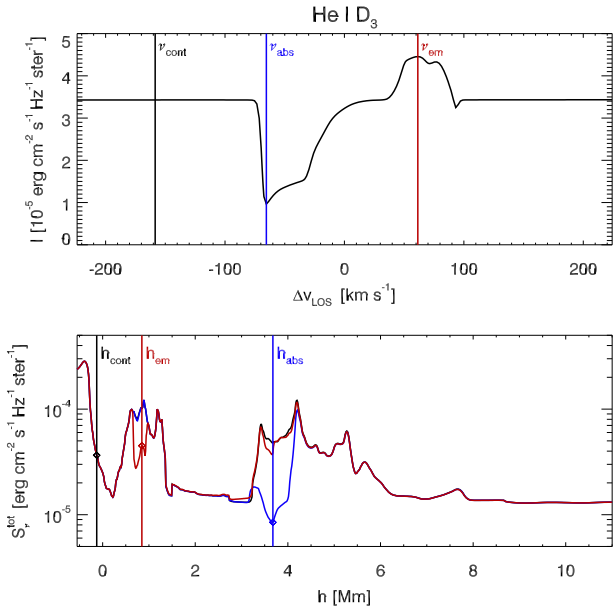}
\caption{A comparison between the emergent intensity and the source function of an EB \hed emission profile. Top panel: emergent intensity. Three vertical lines in black, blue and red correspond to three selected frequencies $\nu_{\rm cont}$, $\nu_{\rm abs}$ and $\nu_{\rm em}$. Bottom panel: the total source function at three different frequencies $S^{tot}_{\nu_{\rm cont}}$ (black), $S^{tot}_{\nu_{\rm abs}}$ (blue) and $S^{tot}_{\nu_{\rm em}}$ (red). The indications $h_{\rm cont}$, $h_{\rm abs}$ and $h_{\rm em}$ correspond to the heights at which $\tau (\nu_{\rm cont})$,  $\tau (\nu_{\rm abs})$ and  $\tau (\nu_{\rm em})$ are equal to one. \label{source}}
\end{figure}

Figures \ref{fourpanel} and \ref{fourpanelhei} show diagrams which break down the different factors that make up the contribution function $C_\nu$ to the \hed and \hei line profiles \citepads{1997ApJ...481..500C} 
\begin{equation}\label{contr}
C_\nu=S^{\rm tot}_\nu\cdot\frac{\chi_\nu}{\tau_\nu}\cdot\tau_\nu e^{-\tau_\nu},
\end{equation}
with $\chi_\nu$ the opacity and $\tau_\nu$ the optical depth.
The source function $S^{\rm tot}_\nu$ equals the total source function, which combines the line source function $S^l_\nu$ and the continuum source function $S^c_\nu$ as
\begin{equation}\label{sf}
S^{\rm tot}_\nu=\frac{j^c_\nu+j^l_\nu}{\chi^c_\nu+\chi^l_\nu}=\frac{\eta_\nu S^l+S^c_\nu}{1+\eta_\nu},
\end{equation}
where $\eta_\nu=\frac{\chi^l_\nu}{\chi^c_\nu}$ and $j_\nu$ is the emissivity. The terms can be separated by calculating the line source function $S^{l}$ via 
\begin{equation}\label{linesource}
S^l=\frac{n_uA_{ul}}{n_lB_{lu}-n_uB_{ul}},
\end{equation}
where the subscript $l$ stands for lower level and $u$ for upper level of the transition. $n$ is the population of the level, $A$ is the Einstein coefficient for spontaneous de-excitation and $B$ is the Einstein coefficient for stimulated (de-)excitation.

The contribution function demonstrates that the absorption in the \hed profile is formed in a thin slab at a height of $\sim 4$ Mm while the emission is formed in a thin layer at $\sim  0.8$ Mm, much deeper in the atmosphere. The same is true for \heic, except that the component formed at $\sim 4$ Mm is mostly in emission as well. The line-of-sight velocity has opposite sign at $0.8$ vs. $4$ Mm so that the absorption component gets blueshifted and the emission component gets redshifted. The $\tau_{\nu}=1$ surfaces of both lines demonstrate that the height of the main opacity sources is very frequency dependent because it shoots up from the photosphere straight into the upper chromosphere and back without a smooth transition -- which is also what we expect to see when we scan through images at different frequencies of the \hed and \hei lines. The value of the total source function $S^{\rm tot}_\nu$ as compared to the continuum source function $S^c_\nu$ should be able to indicate whether we expect absorption or emission for a certain contribution at a certain height. Therefore, we provide a line-plot of the source function of \hed at different frequencies with height in Figure \ref{source}. It is clear that the source function at a height of $\sim 0.8$ Mm and between 3 to 4 Mm is lower than the continuum source function at $\sim 0$ Mm. So the emission is present, not because the line source function is enhanced but because the total source function at this height is larger than at $\sim 0$ Mm where the continuum is formed. To understand the real cause of the emission, we have to investigate why the source function and opacity behave as they do.

\begin{figure}
\includegraphics[scale=1]{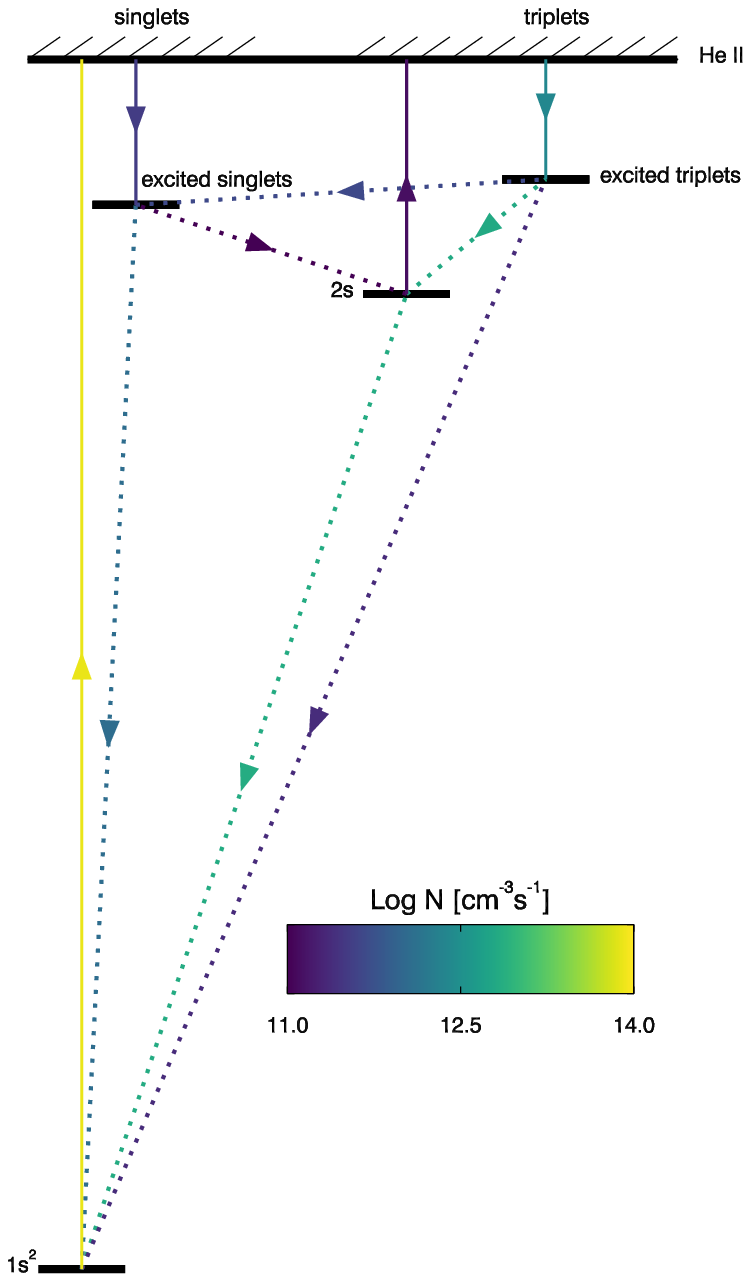}
\caption{Net rate $N$ at location $L_{\rm em}$ (see Fig.~\ref{show_ie}) and height of maximum emissivity in \hed. The model atom incorporates 12 levels in neutral helium (16 in total). We have combined many of these levels to increase the readability of the diagram, but they are treated separately in the calculations (see Sec.~\ref{modelatom} and Fig.~\ref{modelatom}). \label{rate}}
\end{figure}

\begin{figure}
\includegraphics[scale=1]{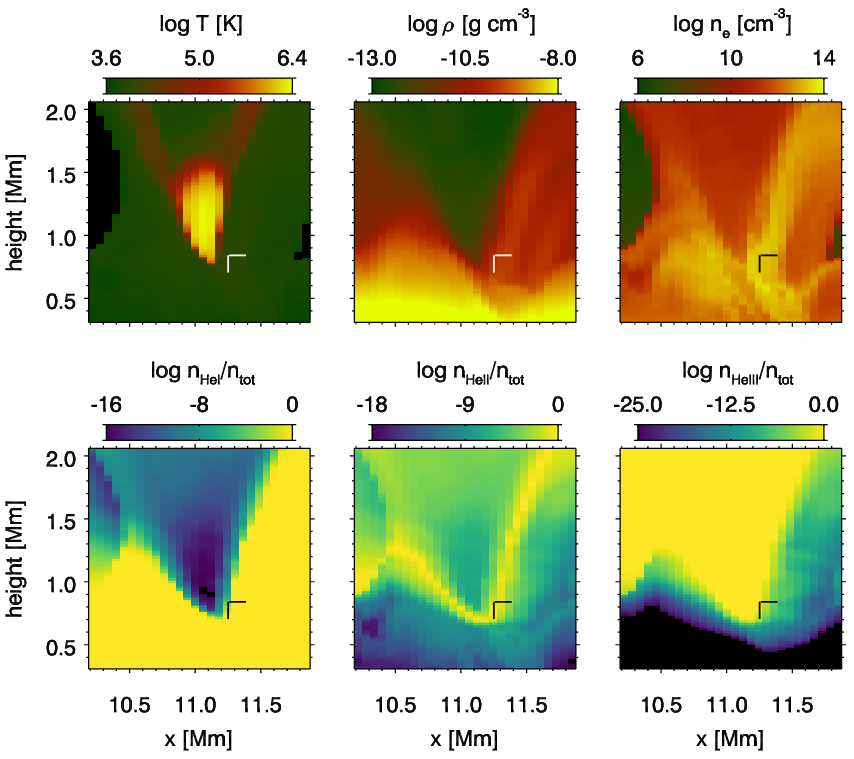}
\caption{Properties of a cut-out along the x-axis of the lower part of the EB (indicated as a white box in Figs.~\ref{vertical_cuts} and \ref{ion_cuts}). Top row: temperature $T$, mass density $\rho$ and electron number density $n_e$ are shown on a log scale. Bottom row: the ratio of the number density of He \textsc{i}, He \textsc{ii} and He \textsc{iii} to the total number density of helium $n_{\rm tot}$ are shown on a log scale. The black and white markers show the position of the \hed profile shown in Figs.~\ref{fourpanel} and \ref{source} at location $\rm L_{em}$ (see Fig.~\ref{obs_synth_zoom}) and $h_{\rm em}$ (see Fig.~\ref{source}). \label{pars}}
\end{figure}

First of all, we have to understand how the triplet levels are populated. Figure~\ref{rate} shows a schematic rate diagram where we have plotted the total net rates $N$, defined as 
\begin{eqnarray}
N & = & n_l(R_{lu}+C_{lu})-n_u(R_{ul}+C_{ul})\\
  & = &( n_lR_{lu}-n_uR_{ul})+(n_lC_{lu}-n_uC_{ul})\\
  & \equiv & N_r + N_c.
 \end{eqnarray}
$R$ are the radiative rates and $C$ the collisional rates between levels $l$ and $u$ or $u$ and $l$. We define $N_r$ and $N_c$ as the total radiative net rate and the total collisional net rate respectively. We have combined many of the levels and net rates accordingly, for a clear visualization. If $N_r>N_c$, the arrow in Fig.~\ref{rate} is plotted as a full line and if $N_c>N_r$ the arrow is plotted as a dotted line. However, the color of the line still represents the total net rate $N$, taking into account both radiative and collisional transitions. 

From Fig.~\ref{rate}, we deduce that the triplet levels are populated by photoionization-recombination. The total number of transitions from the He \textsc{i} ground level to the He \textsc{ii} continuum equals $N=8 \cdot 10^{13}$ cm\textsuperscript{--3} s\textsuperscript{--1}, with $N\sim N_r$ and $N_c=-8 \cdot 10^{8}$cm\textsuperscript{--3} s\textsuperscript{--1} (note the opposite sign). This means there are a factor $10^5$ more photoionizations than collisional recombinations.

Recombination from He~\textsc{ii} into the excited triplet system of neutral helium is also radiatively dominated while the net collisional rates are a factor 10 times smaller and in the opposite direction. Note that there are radiative ionizations from the ground level of the triplet state (indicated as 2s in Fig.~\ref{rate}) balancing the populations in the triplet system. There are no collisions directly populating the triplet system from the ground level of He \textsc{i}, instead a relatively strong downward collisional rate is obtained. All these properties prove that photoionization-recombination is the dominant populating mechanism for the triplet system of He~\textsc{i} at this location. 

%\begin{figure}
%\includegraphics[scale=1]{panels_lines_small}
%\caption{Properties of a cut-out along the x-axis of the lower part of the EB (indicated as a white box in Figs.~\ref{vertical_cuts} and \ref{ion_cuts}). Top left to right: radiation field in  $J_{\nu}$ of \hed at frequency $\nu_{\rm em}$ (see Fig.~\ref{source}), opacity $\chi_\nu$ at $\nu_{\rm em}$ and the line source function of \hed, divided by the local continuum. Bottom left to right: same as top but for \heic. The red markers indicate the position of the \hed profile shown in Figs.~\ref{fourpanel} and \ref{source} at location $\rm L_{\rm em}$ and at $h_{\rm em}$ \label{lines}}
%\end{figure}

Next, we investigate the source of those photoionizing photons at the EB location. We therefore look at a close up of a vertical cut-out along the $x$-axis corresponding to the boxes indicated in Figs.~\ref{vertical_cuts} and \ref{ion_cuts}. The cut-out is shown in Fig.~\ref{pars} at a height of $0.3-2$ Mm. This is the region where \hed emission is formed: $h_{\rm em}=0.8$ Mm in Fig.~\ref{source}. Figure \ref{pars} shows that the very hot part of the bomb with a temperature of $T>10^6$K is located at heights ranging between $0.8$ and $1.5$ Mm, in which all helium is doubly ionized, so only He \textsc{iii} is present. The mass and electron density are $\sim$3 orders of magnitude lower in the high temperature region of the EB compared to outside the EB. Outside the EB, mass and electron density are quite high and increasing quickly with depth because we are situated deep in the atmosphere, only $0.8$ Mm above the $\tau_{500}=1$ surface.
\begin{figure}
\includegraphics[scale=1]{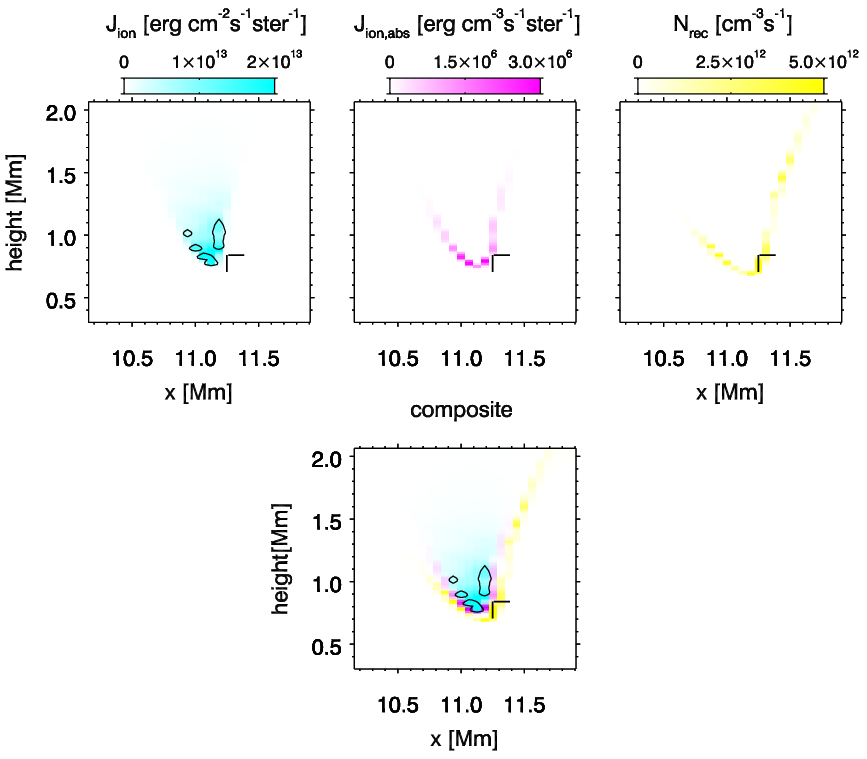}
\caption{Properties of a cut-out along the x-axis of the lower part of the EB (indicated as a white box in Figs.~\ref{vertical_cuts} and \ref{ion_cuts}). Top left to right: the frequency integrated ionizing radiation field $J_{\rm ion}$(see Eq.~\ref{J}), where it is absorbed $J_{\rm ion,abs}$ (see Eq.~\ref{Jabs}) and total number of recombinations in the triplet system $N_{\rm rec}$ as defined Eq.~\ref{nrec}. Bottom: a composite panel of $J_{\rm ion}$, $J_{\rm ion,abs}$ and $N_{\rm rec}$. The black markers indicate the position of the \hed profile shown in Figs.~\ref{fourpanel} and \ref{source} at location at location $\rm L_{em}$ (see Fig.~\ref{obs_synth_zoom}) and $h_{\rm em}$ (see Fig.~\ref{source}).\label{radpans}}
\end{figure}

\begin{figure}
\includegraphics[scale=1]{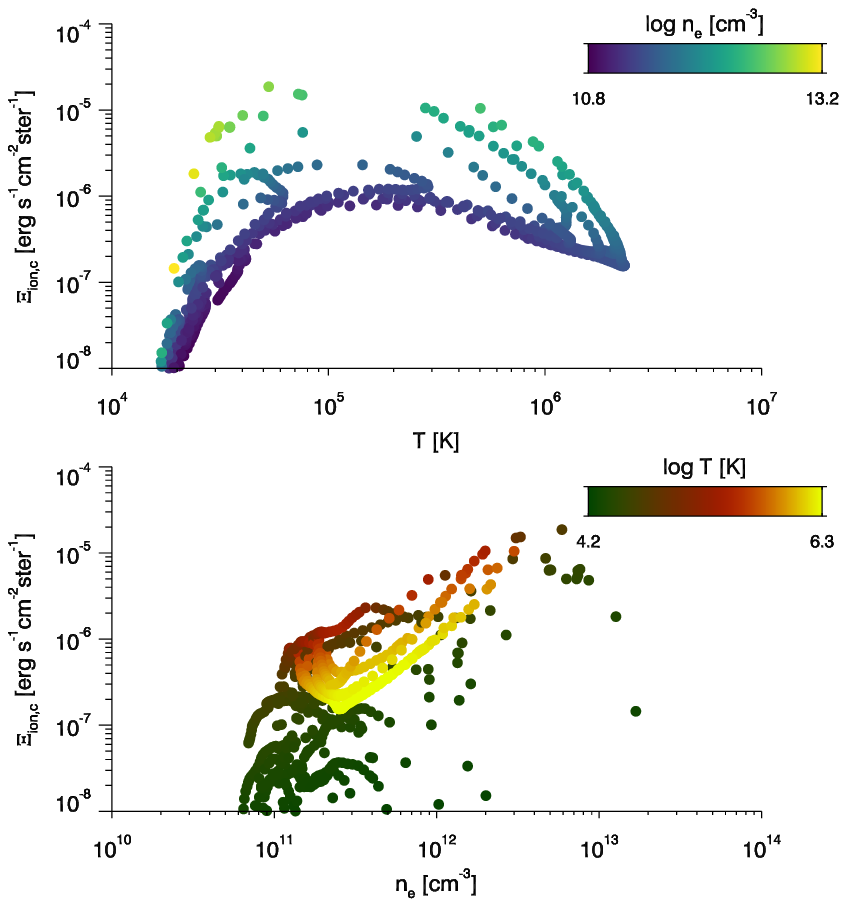}
\caption{Top: all pixels in Fig.~\ref{radpans} in which $\Theta^c_{\rm ion}>10^{-8}\;\rm erg\,s^{-1} cm^{-3} ster^{-1}$ plotted as a function of the temperature in that pixel. The color indicates the electron density in the same pixel. Bottom: all pixels in Fig.~\ref{radpans} in which $\Theta^c_{\rm ion}>10^{-8}\;\rm erg\,s^{-1} cm^{-3}$  plotted as a function of electron density in that pixel. The color indicates the temperature in the same pixel. \label{tempdens}}
\end{figure}

Figure \ref{pars} indicates that a narrow shell is present around the EB in which He \textsc{ii} is present. As it turns out, this shell is vitally important to the formation of the \hed and \hei lines. We know that photoionization-recombination is the mechanism that populates the triplet system of He \textsc{i}. Therefore, in Fig.~\ref{radpans}, we display the EUV radiation field consisting of ionizing radiation integrated between $\nu=0$ and $\nu_{\rm ion}=\frac{c}{\lambda_{\rm ion}}$ with $\lambda_{\rm ion}=504$ \AA\ being the ionization edge of He \textsc{ii}: 
\begin{equation}\label{J}
J_{\rm ion}=\int_\infty^{\nu_{\rm ion}}J_{\nu}\der\nu.
\end{equation}
The emissivity $j_{\rm ion}$ of this radiation can be written as the sum of the background emissivity $j^c_{\rm ion}$ and the emissivity originating from EUV helium lines themselves $j^l_{\rm ion}$:
\begin{equation}
j_{\rm ion} = \int_\infty^{\nu_{\rm ion}}( j^l_{\nu} + j^c_{\nu} )\der\nu = j^l_{\rm ion} + j^c_{\rm ion}.
\end{equation}
The background emissivity consists of two contributions: thermal photons created according to the Planck function $B_{\nu}(T)$ and scattered photons
\begin{eqnarray}\label{jem}
j^c_{\rm ion} & = & \int_\infty ^{\nu_{\rm ion}} \chi^c_{\nu} \left( \epsilon ^c_{\nu}B_{\nu}(T) + (1-\epsilon ^c)J_{\nu}\right)\der\nu \\
            & = & \Theta^c_{\rm ion}+ \Sigma^c_{\rm ion},
\end{eqnarray}
where $\epsilon^c$ is the destruction probability of continuum photons, $\Theta^c_{\rm ion}$ is defined as the thermal contribution to the continuum emissivity, and $\Sigma^c_{\rm ion}$ is the scattering contribution. 
The quantity $J_{\rm ion}$ is shown in Fig.~\ref{radpans} shows us where the ionizing radiation is present while the contours in Fig.~\ref{radpans} tell us where the photons are thermally created, since the contours map the regions where 
\begin{equation}
\Theta^c_{\rm ion}>10^{-6} \,\rm erg\,s^{-1}\,cm^{-3}\,ster^{-1}.
\end{equation}
The ionizing background photons are thermally created close to the lower boundary of the EB. Subsequently, these ionizing photons can propagate only within the hot part of the EB and are absorbed in a shell around the EB, as depicted by the quantity $J_{\rm ion, abs}$ in Fig.~\ref{radpans}, defined as
\begin{equation}\label{Jabs}
J_{\rm ion,abs}=\int_\infty^{\nu_{\rm ion}}J_{\nu}\chi_\nu \der\nu.
\end{equation}
The recombination rate into the triplet system is calculated as 
\begin{equation}\label{nrec}
N_{\rm rec}=n_c\sum_i^{\rm triplet}R_{ci},
\end{equation}
and is proportional to the population of the He \textsc{ii} continuum $n_c$, which peaks in a thin shell around the  EB and around the shell in which $J_{\rm ion,abs}$ peaks.

Since the location where $\Theta^c_{\rm ion}$ peaks is exactly known, we can derive the temperature and electron density required to emit this radiation and hence populate the levels. We selected pixels where $\Theta^c_{\rm ion}>10^{-8} \,\rm erg\,s^{-1}\,cm^{-3}\,ster^{-1}$ to display in Fig.~\ref{tempdens}. The value of $\Theta^c_{\rm ion}$ peaks in regions that have a temperature between $2\cdot 10^4 - 2 \cdot 10^6$ K and with an electron density between $10^{11}$ and $10^{13}\; \rm cm^{-3}$. Data points with the highest $\Theta^c_{\rm ion}$ can have very different temperatures, but all have high electron densities of $10^{13}\; \rm cm^{-3}$.

\begin{figure}
\includegraphics[scale=1]{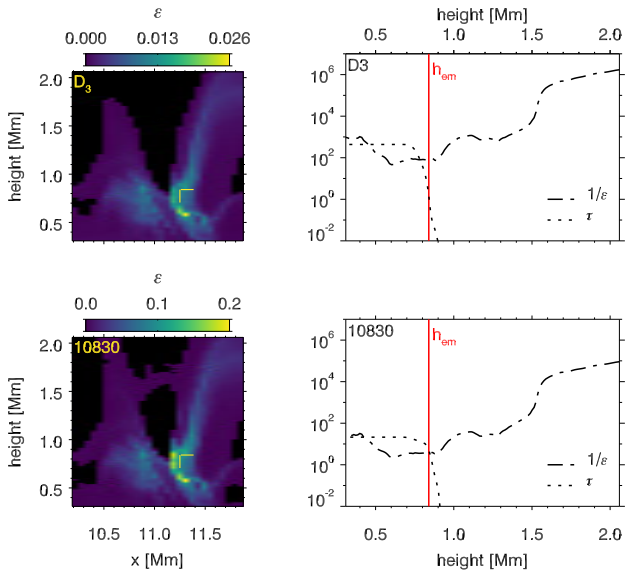}
\caption{Properties of a cut-out along the x-axis of the lower part of the EB (indicated as a white box in Figs.~\ref{vertical_cuts} and \ref{ion_cuts}). Top left: destruction probability $\epsilon$ of \hedc. Bottom left: same for \hei. Top right: thermalization length $1/\epsilon$ (see Eq.~\ref{eqdestr}) in units of optical depth as a function of height at location $L_{\rm em}$ (see Fig.~\ref{obs_synth_zoom}), plotted together with the optical depth $\tau_{\nu}$ at $\nu_{\rm em}$ (see Fig.~\ref{source}). $h_{\rm em}$ as defined in Fig.~\ref{source} is overplotted as a vertical line Bottom right: same for \heic. \label{destr}}
\end{figure}

Once the triplet levels are populated, we still need to study how the lines are formed exactly and why the emission is present. In order to figure out whether thermal processes play a role, it is useful to look at the photon destruction probability with a two-level atom approximation: 
\begin{equation}\label{eqdestr}
\epsilon=\frac{C_{ul}}{C_{ul}+A_{ul}+B_{ul}B(T)},
\end{equation}
where $B(T)$ equals the Planck function. In the two level atom approximation, the source function then consist of a thermal and a scattering part, quantified by $\epsilon$:
\begin{equation}\label{sourceplanck}
S^l=\epsilon B(T) + (1-\epsilon)\bar{J},
\end{equation}
with $B(T)$ the Planck function and $\bar{J}$ equals
\begin{equation}
\bar{J}=\int_0^\infty J_{\nu}\phi(\nu-\nu_0)\der\nu,
\end{equation}
where $\phi(\nu-\nu_0)$ is the line profile and $\nu_0$ the line core frequency.
The photon destruction probability $\epsilon$ is shown in Fig.~\ref{destr} and is a measure for collisional coupling to the local conditions in the plasma, or in other words, how close the conditions are to LTE. It follows roughly the variation of the electron density. The value for $\epsilon$ equals $10^{-2}$ for \hed and $10^{-1}$ for \hei at $L_{\rm em}$ (see Fig.~\ref{obs_synth_zoom}) and $h_{\rm em}$ (see Fig.~\ref{source}). These values are relatively high for chromospheric lines: we can compare with Figs. 12--15 of \citet{johanspaper}, where they show that most chromospheric diagnostics only reach these values in flare ribbons. 

It is useful to compare the thermalization length $1/\epsilon$ in units of optical depth with the optical depth $\tau_\nu$, which is shown in the right panels of Fig.~\ref{destr}. The photons are thermalized when $\tau_\nu \geq 1/\epsilon$. We read from Fig.~\ref{destr} that the opacity of the shell around the EB is optically thick in \hed and \heic. The photons subsequently escape at $h_{\rm em}$ when $\tau_\nu$ drops below one. The temperature in this optically thick layer is in the range $7\cdot 10^3 - 10^4$ K. This is higher than the temperature at which the continuum radiation is formed, and therefore emission is produced due to thermalization. The photons escape subsequently outside the optically thick thermalized part of the layer where $\tau_{\nu}$ drops below one.

Another possible mechanism that could generate emission are recombination cascades in the triplet system. These are however difficult to quantify since the two-level approximation is not valid anymore and one should rewrite Eq.~\ref{sourceplanck} in the following way (see e.g.~\citeads{2012A&A...540A..86R}):
\begin{equation}
S^l=\epsilon' B(T) + (1-\epsilon'-\delta)\bar{J}+\delta B(T').
\end{equation}
Here, $\epsilon'$ is the photon destruction probability but cannot be easily determined and Eq.~\ref{eqdestr} does not apply in this case, $\delta$ is the fraction of photons that are created via alternative processes -- for example recombination cascades -- and $B(T')$ is the Planck function calculated using a virtual temperature $T'$ which corresponds to the radiation temperature of the photons created via recombination cascades. For \heic, we are quite certain that recombination cascades are negligible: the line source function coincides with the temperatures in the thermalized shell because of the high value of $\epsilon$. This strongly suggests that thermal processes are dominating. For \hedc, $\epsilon$ is one order of magnitude smaller and the temperature and source function are only partially coupled. So it could be possible that electron cascades are contributing to the formation of emission in \hedc, but likely not dominating.

\section{Discussion}
The first question that we should address is whether the rMHD simulation presented in \citetads{Hansteen} provides us with a realistic model of an EB. On the positive side, the physics that are associated with reconnection in EBs are present: colliding vertical magnetic fields and resulting bi-directional jets. Also the main spectral characteristics of EBs/UV bursts are reproduced in H$\alpha$, Si \textsc{iv} 1400 \AA\, and Ca \textsc{ii} 8542 \AA. The least realistic part of the simulation in the context of our calculations, is that the coronal radiative losses have been treated in 1D. Therefore, a strongly heated column is present right above the EB starting at a height of $\sim 4$ Mm extending upwards throughout the domain. If coronal losses would have been treated in 3D, these would have propagated in all directions which would likely have a tempering effect. However, we have focused on the emission in \hed generated in the deep atmosphere at $\leq 1$ Mm. Therefore, the emission is not affected by the heated column located above the EB. The \hed and \hei profiles that are shown in the middle column of Fig.~\ref{obs_synth_zoom} at a heliocentric angle of $\mu=0.27$ are unaffected by the heated column, since the line of sight does not cross this region of the domain.
%However, absorption part of the profile shown in Figs.~\ref{fourpanel} and \ref{source} is formed at the bottom boundary of the heated column at 4 Mm. Likely, the transition from chromospheric to coronal temperatures would have occurred higher in the atmosphere and hence the absorption part of the \hed profile would have been formed higher in the atmosphere. The \hed and \hei profiles that are shown in the middle column of Fig.~\ref{obs_synth_zoom} at a heliocentric angle of $\mu=0.27$ are unaffected by the heated column, since the line of sight does not cross this region of the domain.

The largest difference between the observed profiles and the synthetic ones is that the emission peaks are much more intense in the synthetic profiles. The most probable cause of this is the massive amount of material at chromospheric temperatures in this snapshot. In more quiet-sun-like environments, the chromosphere is a rather thin layer usually situated between heights of $\sim 0.5$ and 2 Mm. In this case, the violent flux emergence has pushed up the chromosphere and transition region to heights of up to 8 Mm, as displayed in Fig.~\ref{vertical_cuts}. Realistic or not, this property of the synthetic atmosphere does not affect the physical mechanisms that we propose in this paper, populating the helium triplet levels and generating emission in \hed and \hei.

Another question that we should address is whether it is realistic that the EB/UV burst in the rMHD snapshot reaches temperatures of $10^6$ K. Part of our motivation for this paper originates from discussions and uncertainties about the temperature of EBs, but no estimate in the literature so far has been higher than $10^5$~K. However, these temperatures have always been estimated using chromospheric and transition region spectral lines such as H$\alpha$ and Si \textsc{iv} 1400 \AA. None of these lines would sample plasma with coronal temperatures, since these species get fully ionized. Coronal emission from the EB would on the other hand be completely absorbed by the overlying cooler plasma. This means that we are observationally blind to this type of temperatures in the deep atmosphere and that we have no straightforward way of checking whether it is possible to reach coronal temperatures in EBs. 

Our results show that the ionizing radiation is thermally created at temperatures in the range of $2\cdot 10^4 - 10^6$ K and densities between $10^{11}$ and $10^{13}\,\rm cm^{-1}$. This means that at least in our simulation, we require high temperature in order to generate opacity in \hed and \hei via the ionization-recombination mechanism. This ionizing radiation is created locally in the Ellerman bomb and ionizes a cooler layer around it. This is reminiscent of \cite{2016A&A...594A.104L}, who showed that a substation part of the \hei opacity is created locally adjacent to hot transition region patches. 

Once the triplet levels in neutral helium are populated, the \hei and \hed emission is created via thermalization in an optically thick layer around the EB with temperatures between $7\cdot 10^3$ and $10^4$ K and electron densities of the order of $10^{13}$ cm\textsuperscript{--3}. There might be a possible but not dominant contribution from recombination cascades to the \hed emission.

It is clear that EBs and UV bursts are complex multi-thermal objects. We can think of it as an onion with a hot core and shells around it in which ionizing radiation is created, absorbed and recombinations into the triplet system take place. In \citetads{2017A&A...598A..33L}, we have given a temperature estimate of EBs in the range of $2\cdot 10^4 - 10^5$ K. However, we estimated the lower limit of $2\cdot 10^4$ K based on the assumption of LTE, not based on the idea of generating local ionizing EUV radiation. In any case, we claim -- similarly as in \citetads{2017A&A...598A..33L} -- that the presence of \hed and \hei emission is an indicator of high temperatures in EBs. This is compatible with the radiative transfer calculations presented by \citetads{Hansteen}, which confirm that even an EB/UV burst of $10^6$ K can be compatible with the observed H$\alpha$ and Si \textsc{iv} 1400 \AA\ profiles in these objects.

\section{Summary and conclusion}
In this paper, we have focused on helium line formation in an EB. We made use of a synthetic atmosphere generated by an rMHD simulation and we used the atmosphere to calculate helium line intensities using the 3D non-LTE code Multi3D.
 
The EB line profiles of \hei and \hed in the simulation show similar properties as the observed line profiles of an EB at equal viewing angle. Emission in the wing is present while the line core is in absorption. The dynamics an the broadening are largely reproduced in the simulation. However, the line intensities of the simulation are different: both the emission and the absorption signatures are stronger in the synthetic profiles as compared to the observed profiles, at least for the \hed line. Also, the synthetic emission in \hei is very intense while barely present in the observed profiles. 

Encouraged by the qualitative overlap between the synthetic and observed EB \hed profiles, we have continued the paper to focus on the most intense \hed emission profile.
 
The emission part of the \hed profile in our simulation is formed low in the atmosphere at a height of $\sim$ 1 Mm while the absorption part is formed much higher at $\sim$ 4 Mm, reminiscent of the formation of H$\alpha$ profiles in EBs. We find that photoionization-recombination is the dominant mechanism to populate the helium triplet system both inside and outside the EB. The ionizing radiation is thermally created locally inside the EB at temperatures between $2\cdot 10^4$ and $10^6$ K and electron densities ranging between $10^{11}$ and $10^{13}\,\rm cm^{-1}$, and absorbed in a dense and cooler shell around it. Recombinations are populating the triplet levels. Emission in \hed and \hei is created in a thermalized shell which is optically thick in \hed and \hei, and has temperatures in the range of $7\cdot 10^3-10^4$ K and electron densities between $10^{12}$ and $10^{13}\;\rm cm^{-3}$.

These results suggest that the observation of \hed and \hei signatures in EBs/UV bursts indicate temperatures higher than $2\cdot 10^4$ K and in our case as high as $\sim 10^6$ K.

\begin{acknowledgements}
TL is supported by the Swedish Research Council
(2015-03994) and the Swedish National Space Board (128/15). JdlCR is supported by grants from the Swedish Research Council
(2015-03994), the Swedish National Space Board (128/15) and the Swedish
Civil Contingencies Agency (MSB). This project has received funding from
the European Research Council (ERC) under the European Union's Horizon
2020 research and innovation programme (SUNMAG, grant agreement 759548).
JJ is supported by the Research Council of Norway, project 250810, and through its Centres of Excellence scheme, project number 262622.
The Institute for Solar Physics is supported by a grant for research
infrastructures of national importance from the Swedish Research Council
(registration number 2017-00625).

The Swedish 1-m Solar Telescope is operated on the island of La Palma
by the Institute for Solar Physics of Stockholm University in the
Spanish Observatorio del Roque de los Muchachos of the Instituto de
Astrof\'isica de Canarias. 

The 3D non-LTE radiative transfer calculations were performed on resources provided by the Swedish National Infrastructure for Computing (SNIC) at the High Performance Computing Center North at Ume\aa~University.

This research has made use of NASA's
Astrophysics Data System Bibliographic Services.
\end{acknowledgements}

\bibliography{references}

\end{document}